\documentclass[conference]{IEEEtran}
\IEEEoverridecommandlockouts
\usepackage{cite}
\usepackage{amsmath,amssymb,amsfonts}
\usepackage{algorithmic}
\usepackage{graphicx}
\usepackage{textcomp}
\usepackage{xcolor}

\usepackage{hyperref}
\usepackage{booktabs} 
\usepackage{makecell}
\usepackage{soul} 

\def\BibTeX{{\rm B\kern-.05em{\sc i\kern-.025em b}\kern-.08em
    T\kern-.1667em\lower.7ex\hbox{E}\kern-.125emX}}
\begin{document}

\title{Does Re-referencing Matter? \\Large Laplacian Filter Optimizes Single-Trial P300 BCI Performance\\
\thanks{This work is supported in part by STI 2030
	Major Projects, the National Key Research and Development Program of China (Grant No. 2022ZD0208500).}
	}
	
	\author{\IEEEauthorblockN{Eva Guttmann-Flury }
\IEEEauthorblockA{
	\textit{Department of Micro-Nano Electronics and } \\
	\textit{the MoE Key Laboratory of Artificial Intelligence} \\
	\textit{Shanghai Jiao Tong University}\\
	Shanghai, China \\
	\textit{and visiting at CenBRAIN Neurotech,} \\
	\textit{School of Engineering, Westlake University}\\
	Hangzhou, China \\
	eva.guttmann.flury@gmail.com}
\and
\IEEEauthorblockN{Jian Zhao}
\IEEEauthorblockA{
	\textit{Department of Micro-Nano} \\
	\textit{Electronics and the MoE Key} \\
	\textit{Laboratory of Artificial Intelligence} \\
	\textit{Shanghai Jiao Tong University}\\
	Shanghai, China \\
	zhaojianycc@sjtu.edu.cn}
\and
\IEEEauthorblockN{Mohamad Sawan}
\IEEEauthorblockA{
	\textit{CenBRAIN Neurotech,} \\
	\textit{School of Engineering} \\
	\textit{Westlake University}\\
	Hangzhou, China \\
	sawan@westlake.edu.cn}

}

\maketitle

\begin{abstract}
Electroencephalography (EEG) provides a non-invasive window into brain activity, enabling Brain-Computer Interfaces (BCIs) for communication and control. However, their performance is limited by signal fidelity issues, among which the choice of re-referencing strategy is a pervasive but often overlooked preprocessing bias. Addressing controversies about its necessity and optimal choice, we adopted a quantified approach to evaluate four strategies -- no re-referencing, Common Average Reference (CAR), small Laplacian, and large Laplacian -- using 62-channels EEG (31 subjects, 2,520 trials). To our knowledge, this is the first study systematically quantifying their impact on single-trial P300 classification accuracy. Our controlled pipeline isolated re-referencing effects for source-space reconstruction (eLORETA with Phase Lag Index) and anatomically constrained classification. The large Laplacian resolves distributed P3b networks while maintaining P3a specificity, achieving the best P300 peak classification accuracy (81.57\% hybrid method; 75.97\% majority regions of interest). Performance follows a consistent and statistically significant hierarchy: large Laplacian $>$ CAR $>$ no re-reference $>$ small Laplacian, providing a foundation for unified methodological evaluation.
\end{abstract}


\section{Introduction}
Electroencephalography (EEG)-based Brain-Computer Interfaces (BCIs) offer transformative potential for clinical neurotechnology and cognitive state monitoring. However, their real-world deployment is constrained by the inherent variability of single-trial neural signatures, which arises from both biological sources (e.g., inter-subject neuroanatomical differences, attentional fluctuations) and technical limitations (e.g., signal contamination, spatial smearing). Such instability severely compromises decoding accuracy, particularly for applications requiring consistent biomarker detection \cite{Maswanganyi_2022, Gonzalez_2021, Roh_2014}.

Within this landscape, the P300 signal stands out as a stable neurophysiological anchor due to its consistent spatiotemporal properties, including its topography, biphasic morphology, and predictable post-stimulus latency, typically occurring within standard time windows (P3a: 150 -- 250 ms; P3b: 300 -- 500 ms). These characteristics exhibit lower intra- and inter-subject variability compared to other neural features, making the P300 an ideal candidate for isolating technical confounds that undermine EEG signal fidelity. Among these potential biases, re-referencing methodology represents a pervasive yet underquantified source of systematic distortion \cite{Yano_2019, Eyamu_2024}.

This study examines four strategies representing complementary approaches to EEG's intrinsic indeterminacy. The first approach preserves the original amplifier reference (located on the right mastoid), maintaining raw signal topology while accepting the limitations of physical electrode references: susceptibility to drift and bioelectric contamination \cite{Delorme_2023}. In contrast, the Common Average Reference (CAR) imposes statistical centralization by subtracting the global mean scalp potential, conceptually approximating a neutral baseline yet relying on the biophysically untenable assumption of head equipotentiality. This assumption inevitably smears local activity through volume conduction effects \cite{Tsuchimoto_2021}.

The small Laplacian transform, embodies a localist philosophy by computing spatial derivatives through nearest-neighbor differentiation. This approach emphasizes cortical sources within a restricted spatial domain, physically manifesting Ohm's law principle where current source density corresponds to the second spatial derivative of electrical potential. While enhancing focal generators, it risks attenuating distributed activity essential for integrated neural processes. Complementarily, the large Laplacian extends differentiation to incorporate broader cortical neighborhoods, capturing source dynamics across functionally integrated regions. This balanced approach better models the patch-like organization of P300 generators while maintaining reasonable spatial specificity \cite{Nunez_2019, Cisotto_2024}.

\begin{figure}[!htp]
	\centering
	\includegraphics[width=7cm]{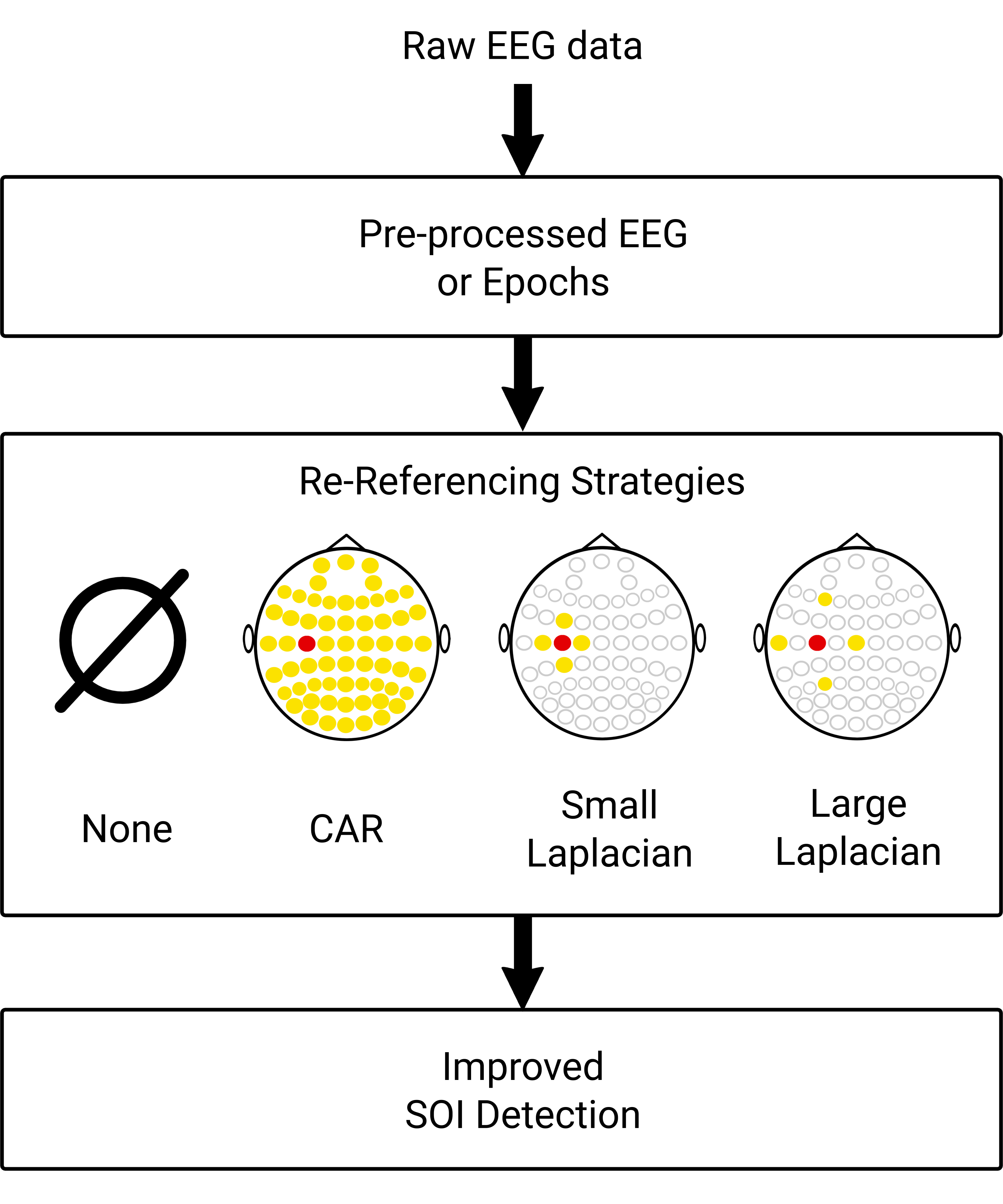}
	\caption
	[Goal]
	{Flowchart for quantifying three different re-referencing strategies: Common Average Reference (CAR), Small Laplacian, and Large Laplacian, with None as baseline}
	\label{fig:Goal}
\end{figure}

This study examines four reference strategies representing distinct approaches: maintaining the original reference reveals inherent recording limitations; CAR applies statistical centralization; while small and large Laplacian transforms estimate current source density at local and extended scales (Fig \ref{fig:Goal}). Selected for their complementary biophysical foundations and methodological prevalence, these approaches present divergent neurofunctional implications for P300 analysis -- CAR may obscure frontoparietal distinctions while Laplacians risk oversegmenting integrated networks. Crucially, the optimal strategy depends on the signal's neurophysiological properties, particularly its effect size: while the robust P300 (with its high SNR and consistent topography) provides a stable testbed, smaller-effect components would exhibit greater vulnerability to reference-induced distortions. To the best of our knowledge, no study has systematically quantified their impact on single-trial P300 classification performance.

This research constitutes a fundamental effort toward a unified evaluation of methodological approaches across various preprocessing parameters. Our hypothesis posits that the choice of re-referencing exerts a higher influence on subsequent BCI performance compared to other pipeline variables. To investigate this, we employ a multifaceted approach integrating biophysical modeling through boundary element methods, neuroanatomical validation using source localization, and single-trial performance metrics. By assessing the impact of each strategy throughout the analytical process, we seek to develop evidence-based guidelines for standardizing preprocessing pipelines, thereby advancing the development of clinically viable BCIs rooted in biophysical principles.

\section{Methods}

\subsection{Experimental Data}
The study employs the Eye-BCI multimodal dataset \cite{eye_bci_multi_dataset}, comprising high-density (62-channel) EEG recordings from 31 healthy adults (age 22-57 years; 29 right-handed) during 2,520 visual P300 trials. This dataset provides comprehensive artifact monitoring through synchronized EOG, EMG, and eye-tracking \cite{GuttmannFlury_Dataset_2025}, along with a statistically optimized sample size determined via Fitted Distribution Monte Carlo (FDMC) power analysis \cite{GuttmannFlury_APriori_2019}. The combination of high temporal resolution (1000 Hz sampling) and rigorous artifact characterization establishes an ideal foundation for investigating re-referencing effects on P300 dynamics. Fully documented experimental protocols are available in a public repository (\href{https://github.com/QinXinlan/EEG-experiment-to-understand-differences-in-blinking/}{https://github.com/QinXinlan/EEG-experiment-to-understand-differences-in-blinking/}). 

\subsection{Channel Validation and Artifact Correction}
The preprocessing pipeline begins with systematic channel validation using the Adaptive Blink and Correction De-drifting (ABCD) algorithm \cite{GuttmannFlury_BadChannel_2025}. This automated procedure identifies compromised electrodes through quantitative comparison of blink-related potentials across spatial neighborhoods. This spatial consistency check ensures robust identification of hardware artifacts and poor contact channels while preserving neurophysiological signals.

The ABCD algorithm then performs simultaneous correction of blink artifacts and low-frequency drift while maintaining the original 1000 Hz temporal resolution. The blink correction employs a template subtraction approach that accounts for spatial variability in ocular artifacts, while the drift removal utilizes adaptive spline fitting to preserve millisecond-scale P300 latency information. This dual correction strategy operates on the principle of minimal signal distortion, ensuring that subsequent re-referencing analyses reflect genuine neural activity rather than artifact remnants.

\subsection{Re-referencing Methodology} 
EEG signals measure potential differences \mbox{$V_{i,ref} = \phi_i-\phi_{ref}$}, between each electrode $\phi_i$ and a reference $\phi_{ref}$, which lacks a physiological zero due to volume conduction and the non-uniqueness of inverse solutions. The choice of reference is deemed critical because it introduces spatial and spectral biases: scalp potentials reflect volume-conducted macro-scale activity (e.g., synchronous dipole layers spanning 10 -- 50 $cm^2$ of cortex), while local micro-scale sources (e.g., synaptic currents) may cancel out or distort depending on reference strategy. Re-referencing aims to mitigate these biases, but controversies persist. Some argue it artificially centralizes data (e.g., CAR), while others highlight its necessity for isolating genuine cortical activity from non-neutral references (e.g., mastoid or nose references contaminated by non-neural noise) \cite{Delorme_2023, SuarezRevelo_2018, Nunez_2019}. 

\paragraph{Common Average Reference}
CAR addresses the reference problem by subtracting the mean potential \mbox{$\overline{\phi} = \frac{1}{N} \sum_{j=1}^{N} \phi_j$} across all electrodes from each individual channel $\phi_{{CAR}}^{(i)} = \phi^{(i)} - \overline{\phi}$. This assumes $\overline{\phi} \approx 0$, valid only for spherical head models with uniformly distributed sources. While CAR suppresses global noise (e.g., volume-conducted activity from distant cortical regions), it distorts local activity in three key scenarios: (1) Incomplete electrode coverage, where missing regions bias the average; (2)~Asymmetric~source distributions, common in non-spherical heads or pathological conditions; and (3) Deep sources, which contribute negligibly to the scalp average but may be critical for EEG analysis \cite{Huster_2010}. Additionally, CAR’s rank deficiency (loss of one degree of freedom) further limits its utility for source localization, as it conflates genuine cortical dipoles with reference artifacts \cite{Nunez_2019}.

\paragraph{Small Laplacian}
The Small Laplacian, or Local Hjorth Laplacian, estimates current source density (CSD) through spatial differentiation of nearest-neighbor potentials: 
\mbox{$L_{\text{small}}^{(i)} = \phi^{(i)} - \frac{1}{|N(i)|} \sum_{j \in N(i)} \phi^{(j)}$}, approximating the second spatial derivative of scalp potentials to emphasize focal cortical generators. This method enhances spatial specificity by suppressing volume-conducted contributions from distant sources, operating 
independently of reference choice -- a critical advantage for isolating local activity. 
The accuracy of this spatial derivative is constrained by electrode density; a minimum of 64 channels is a standard heuristic to mitigate spatial undersampling for uniform, whole-head coverage, ensuring each electrode has a sufficient number of neighbors. However, for a carefully selected, task-relevant subset of electrodes focused over a specific cortical region, this requirement can be significantly lower \cite{Meng_2018, GuttmannFlury_Channel_2022}. The method remains sensitive to noise and boundary effects where incomplete neighbor sets distort estimates.
Crucially, the Small Laplacian attenuates distributed sources spanning $>$ 1 cm, as it intrinsically filters low spatial frequencies associated with large-scale synchrony. This limitation arises because scalp potentials from macro-scale dipole layers are governed by tangential current flows that cancel in nearest-neighbor derivatives, rendering it suboptimal for distributed cognitive processes \cite{Hjorth_1991, Nunez_2006}.

\paragraph{Large Laplacian}
This extended Laplacian  enhances the localization of cortical electrical activity by considering a broader spatial context: \mbox{$L_{\text{large}}^{(i)} = \phi^{(i)} - \frac{1}{|\mathcal{R}_i^{\text{large}}|} \sum_{j \in \mathcal{R}_i^{\text{large}}} \phi^{(j)}$}, where $\mathcal{R}_i^{\text{large}}$ encompasses electrodes within concentric rings spanning 50 -- 100\% of the maximum inter-electrode distance (e.g., F3, T7, P3, and Cz for electrode C3 in standard 10-10 configurations). This formulation establishes a band-pass spatial filter that attenuates both local high-frequency noise through spatial averaging and far-field volume conduction from distant sources, while maximizing sensitivity to cortical patches of 3 -- 6 cm diameter \cite{Nunez_2006}. Physiologically, this scale aligns with mesoscale synchrony in cortical dipole layers, where microsources exhibit moderate spatial coherence -- exceeding minicolumns ($<$ 1 cm) but remaining subordinate to lobar networks ($>$ 10 cm) \cite{Buxhoeveden_2002, Sporns_2007}. 

The core trade-offs of all re-referencing strategies are comparatively summarized in Table~\ref{tab:reref}.
\begin{table}[htbp]
	\centering
	\caption{Re-referencing Strategy Comparison}
	\label{tab:reref}
	\begin{tabular}{ccc}
		\toprule
		\textbf{Method}  & \textbf{Advantages} & \textbf{Limitations} \\
		\midrule
		\makecell{No \\Re-referencing } & \makecell{Preserves raw signal \\topology}    & \makecell{Reference contamination \\biases} \\
		\\
		\makecell{CAR } & \makecell{Reduces whole-head \\noise}    & \makecell{Distorts local activity; \\Smears spatial topography} \\
		\\
		\makecell{Small \\Laplacian } & \makecell{High spatial specificity; \\ Reference-independent}    & \makecell{Attenuates distributed \\activity; \\Noise amplification} \\
		\\
		\makecell{Large \\Laplacian } & \makecell{Optimized for \\mesoscale; \\Balances specificity + \\network integrity}    & \makecell{Reduced focal \\source resolution; \\Requires dense \\electrode arrays} \\
		
		\bottomrule
	\end{tabular}
\end{table}

\subsection{Source-Space Quantification and P300 Classification}
Before source reconstruction, preprocessed signals undergo zero-phase bandpass filtering using a 4th-order Butterworth filter (1 –- 15 Hz) to isolate the P300's characteristic spectral range. Epochs are segmented from $-\,200$ to $+\,700$ ms relative to stimulus onset, retaining pre-stimulus baselines ($-\,200$ -- \mbox{$0$ ms}) for amplitude normalization. This temporal window captures the full spatiotemporal evolution of P300 subcomponents while minimizing edge artifacts.

Source-space analysis then transforms these preprocessed scalp EEG signals into neuroanatomical insights by solving the electromagnetic inverse problem through biophysical modeling. Using the FreeSurfer fsaverage template for cross-subject anatomical standardization, we leveraged MNE-Python's precomputed boundary element forward model with tissue-specific conductivities (scalp: $0.33\,S/m$, skull: $0.006\,S/m$, brain: $0.33\,S/m$) to map sensor-space signals to cortical generators \cite{Gramfort_2013}. Distributed source reconstruction via exact Low-Resolution Electromagnetic Tomography (eLORETA) was applied without a priori assumptions, discretizing the cortex into 10,242 dipoles per hemisphere to resolve spatially overlapping P300 generators while minimizing anatomical variability \cite{PascualMarqui_2018}.

Functional connectivity characterized transient network dynamics underlying P300 subcomponents within specific time windows (P3a: 150 -- 250 ms; P3b: 300 -- 500 ms). The Phase Lag Index (PLI) quantified phase synchronization across 68 cortical parcels, specifically chosen for its capacity to reject volume conduction artifacts by ignoring zero-lag phase interactions. This metric prioritizes consistent non-zero phase lags reflecting true neurofunctional coupling while suppressing spurious correlations from field spread \cite{Stam_2007}. To identify stable regions of interest, we retained only the top 20 connection pairs per session, defining ROIs as the three brain regions most consistently appearing across participants.

To classify single-trial P300 responses, we compute activation maxima within these cross-subject ROIs across the entire epoch ($-\,200$ to $+\,700$ ms). Trials are labeled as P300-positive if their global maxima occur within the neurophysiologically plausible interval ($+\,200$ to $+\,550$ ms), leveraging the anatomical specificity of eLORETA-derived ROIs. To address potential session-specific ROI misalignment, we apply a secondary temporal agreement criterion requiring maxima across a majority of ROIs to cluster within this same latency window. Finally, a hybrid method combines both criteria, accepting trials that satisfy either the spatial maximum condition or the temporal consensus requirement.

\section{Results}

\subsection{Network Topography}
\begin{figure}[!htp]
	\centering
	\includegraphics[width=9cm]{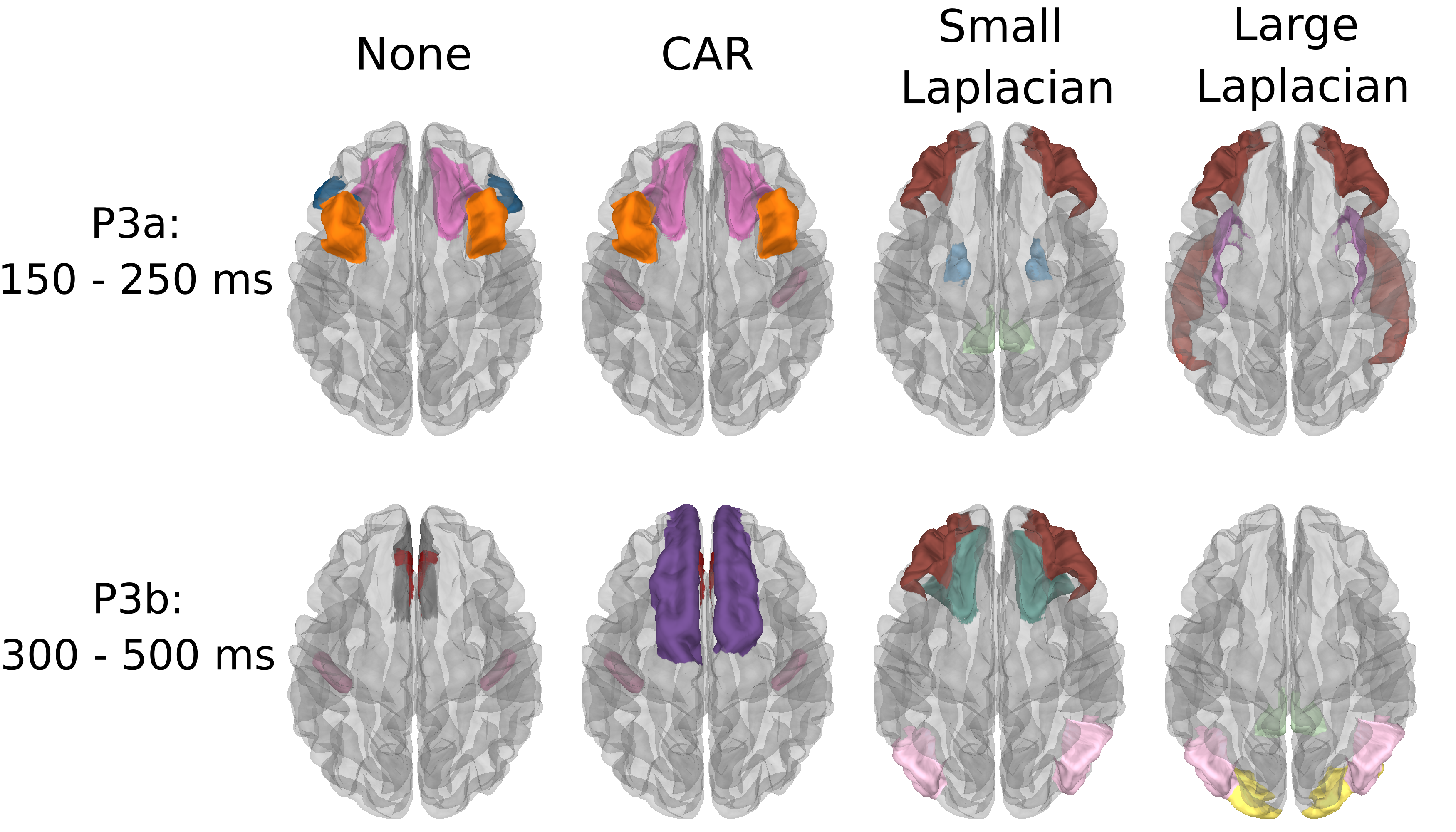}
	\caption
	[P300 Network Topography]
	{Re-referencing strategies reveal distinct P3a/P3b network signatures. The colors represent the different brain regions (consistent across strategies)}
	\label{fig:Topo}
\end{figure}

Connectivity profiles align with the hierarchical neural mechanisms governing visual P300 generation, where distributed networks coordinate attentional capture, sensory integration, and memory updating. PLI analysis consistently identified fronto-insular-parietal hubs across all re-referencing conditions, with rostral middle frontal and insular regions driving attentional orienting (P3a) through delta-theta synchronization. Crucially, the large Laplacian strategy demonstrated superior resolution of the full neuroanatomical hierarchy, particularly capturing middle temporal and inferior parietal alpha-theta interactions underlying context updating (P3b) with enhanced spatial fidelity (Fig. \ref{fig:Topo}).

This contrasts with other strategies: CAR exhibited spatial smearing that attenuated posterior hub specificity, the small Laplacian overemphasized frontal generators while suppressing distributed P3b networks, and the original reference introduced anterior bias through volume-conducted noise. The large Laplacian's balanced sensitivity to both focal and distributed sources -- especially deeper temporoparietal generators -- validates its theoretical advantage for mesoscale cortical dynamics.

\subsection{Statistical Validation}
Single-trial P300 classification accuracies demonstrate statistically significant differences across re-referencing strategies.For each strategy, we first computed the mean accuracy $\mu$ and standard deviation across all sessions (N = 63). The 95\% confidence intervals for these means are then derived as $CI = \mu \pm Z \cdot \frac{SD}{\sqrt{N}}$, using the standard Normal Z-score $Z = 1.96$. To quantify the degree of separation between these point estimates, we employ a novel normalized confidence interval divergence metric:
\begin{equation}
	\Delta_{A,B} = \frac{ \mathcal{LB} ({\text{Max}_{A,B}}) - \mathcal{UB} ({\text{Min}_{A,B}}) }{ \left| \mu_A - \mu_B \right|}
\end{equation}
where $\mu$ represents mean accuracy, and $\mathcal{LB}$ and $\mathcal{UB}$ denote lower and upper 95\% confidence bounds, computed as ${\text{Max}_{A,B}} - \frac{1}{2} {\text{CI}}$ (resp. ${\text{Min}_{A,B}} + \frac{1}{2} {\text{CI}}$). A $\Delta_{A,B} > 1$ indicates non-overlapping confidence intervals under normality assumptions, confirming statistically distinct performance between strategies $A$ and $B$.

\begin{figure}[!htp]
	\centering
	\includegraphics[width=9cm]{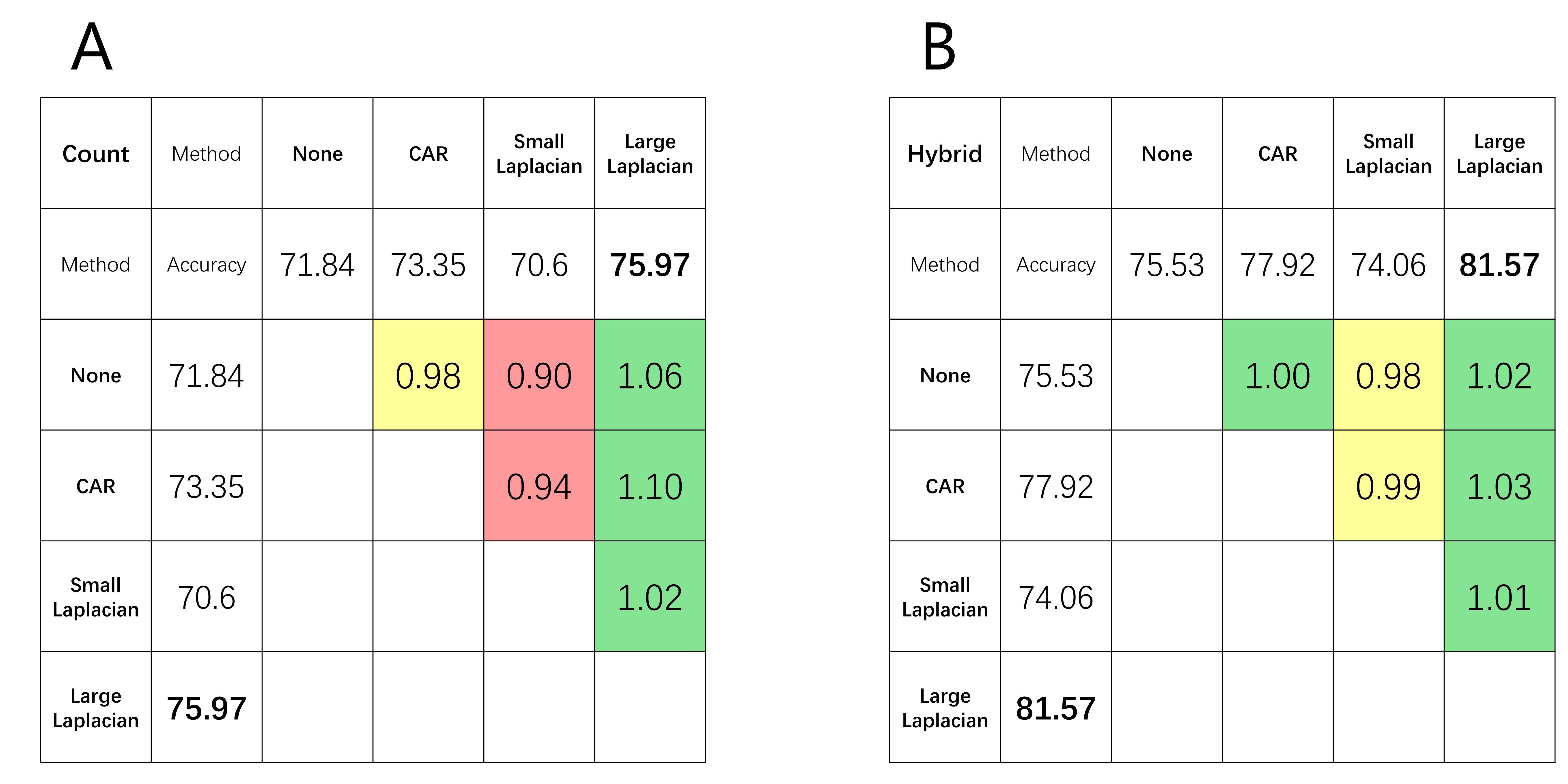}
	\caption
	[Classification]
	{Pairwise performance comparison matrices of re-referencing strategies for single-trial P300 classification based on the normalized confidence interval divergence for the (A)~majority-ROI and (B) hybrid methods. Green~($\Delta_{A,B}~>~1$): significant; Yellow~($0.95 < \Delta_{A,B} \leq 1$): marginal; Red~($\Delta_{A,B} \leq 0.95$): non-significant.}
	\label{fig:Classif}
\end{figure}

This divergence metric analysis reveals a consistent performance hierarchy across re-referencing strategies for the single-trial P300 classification. The large Laplacian achieves superior decoding fidelity, followed by CAR, with original reference and small Laplacian showing progressively reduced performance. All pairwise comparisons ($\Delta_{A,B}$) are illustrate in Fig. \ref{fig:Classif}. The color legend illustrates non-overlapping CIs in green, minimally overlapping CIs in yellow, and overlapping CIs (i.e., non-significant difference) in red.

\section{Discussion}

This study establishes that re-referencing strategy fundamentally determines BCI performance through its cascading effects across the analytical pipeline. Our integrated approach reveals that large Laplacian processing optimally preserves P300 neurophysiology, achieving peak classification accuracies of 75.97\% (majority-count ROIs) and 81.57\% (hybrid method) -- significantly outperforming alternative strategies. Although this work was carried out on a single dataset, these findings align with and extend previous research on re-referencing strategies \cite{McFarland_2015, Liu_2021}, specifically confirming the P300's sensitivity to spatial filtering scales.

Methodologically, this work provides a rigorous template for EEG pipeline optimization by systematically isolating a single preprocessing parameter while controlling all other variables. Future studies should apply this same controlled approach to evaluate other critical parameters: the number of ROIs retained for classification, source localization algorithms, and head model specifications all represent potential optimization targets that may interact with re-referencing choices.

This paper represents a foundational step toward developing integrated methodologies that rigorously evaluate each step of the EEG processing chain. By establishing this systematic evaluation framework and demonstrating its utility through quantified trade-offs, we provide both immediate guidance for P300-based BCIs and a template for comprehensive pipeline optimization across neurotechnology applications.

\section{Conclusion}

This work provides the first step towards a quantitative resolution to EEG re-referencing controversies through a novel methodology that integrates neurophysiological validation with applied performance metrics. Our controlled pipeline demonstrates that reference choice helps reconstruct brain dynamics and determines BCI efficiency, establishing large Laplacian as optimal for the P300 signal of interest. This approach creates a generalizable gold standard for preprocessing optimization, offering immediate value for clinical neurotechnology and transferable principles for reliable real-world BCIs.



\begin{thebibliography}{00}

\bibitem{Maswanganyi_2022} R. C. Maswanganyi, C. Tu, P. A. Owolawi and S. Du, ``Statistical evaluation of factors influencing inter-session and inter-subject variability in eeg-based brain computer interface,'' in IEEE Access, vol. 10, pp. 96821--96839, September  2022.  \href{https://doi.org/10.1109/ACCESS.2022.3205734}{https://doi.org/10.1109/ACCESS.2022.3205734}


\bibitem{Gonzalez_2021} H. A. Gonzalez, R. George, S. Muzaffar, J. Acevedo, S. Hoeppner, C. Mayr, J. Yoo F. H. P. Fitzek  and I. M.  Elfadel, ``Hardware acceleration of EEG-based emotion classification systems: a comprehensive survey,'' in IEEE Transactions on Biomedical Circuits and Systems, vol. 15, no. 3, pp. 412--442, June 2021.  \href{https://doi.org/10.1109/TBCAS.2021.3089132}{https://doi.org/10.1109/TBCAS.2021.3089132}

\bibitem{Roh_2014} T. Roh, K. Song, H. Cho, D. Shin and H. -J. Yoo, ``A Wearable Neuro-Feedback System With EEG-Based Mental Status Monitoring and Transcranial Electrical Stimulation,'' in IEEE Transactions on Biomedical Circuits and Systems, vol. 8, no. 6, pp. 755-764, December 2014.  \href{https://doi.org/10.1109/TBCAS.2021.3089132}{https://doi.org/10.1109/TBCAS.2021.3089132}

\bibitem{Yano_2019} M. Yano, S. Suwazono, H. Arao, D. Yasunaga and H. Oishi, ``Inter-participant variabilities and sample sizes in P300 and P600,'' in International Journal of Psychophysiology, vol. 140, pp. 33--40, 2019.  \href{https://doi.org/10.1016/j.ijpsycho.2019.03.010}{https://doi.org/10.1016/j.ijpsycho.2019.03.010}

\bibitem{Eyamu_2024} J. Eyamu, W. S. Kim, K. Kim, K. H. Lee and J. U. Kim, ``Prefrontal intra-individual ERP variability and its asymmetry: exploring its biomarker potential in mild cognitive impairment,'' in Alzheimer's Research \& Therapy, vol. 16, no. 1, pp. 83, 2024.  \href{https://doi.org/10.1186/s13195-024-01452-5}{https://doi.org/10.1186/s13195-024-01452-5}

\bibitem{Delorme_2023} A. Delorme, ``EEG is better left alone,'' in Scientific reports, vol. 13, no. 1, pp. 2372, 2023.  \href{https://doi.org/10.1038/s41598-023-27528-0}{https://doi.org/10.1038/s41598-023-27528-0}

\bibitem{Tsuchimoto_2021} S. Tsuchimoto, S. Shibusawa, S. Iwama, M. Hayashi, K. Okuyama, N. Mizuguchi, , K. Kato and J. Ushiba, ``Use of common average reference and large-Laplacian spatial-filters enhances EEG signal-to-noise ratios in intrinsic sensorimotor activity,'' in Journal of Neuroscience Methods, vol. 353, pp. 109089, April 2021.  \href{https://doi.org/10.1016/j.jneumeth.2021.109089}{https://doi.org/10.1016/j.jneumeth.2021.109089}

\bibitem{Chamanzar_2017} A. Chamanzar, M. Shabany, A. Malekmohammadi and S. Mohammadinejad, ``Efficient hardware implementation of real-time low-power movement intention detector system using FFT and adaptive wavelet transform,'' in IEEE transactions on biomedical circuits and systems, vol. 11, no. 3, pp. 585--596, June 2017.  \href{https://doi.org/10.1109/TBCAS.2017.2669911}{https://doi.org/10.1109/TBCAS.2017.2669911}

\bibitem{Huster_2010} R. J. Huster, R. Westerhausen, C. Pantev and C. Konrad,  ``The role of the cingulate cortex as neural generator of the N200 and P300 in a tactile response inhibition task,'' in Human brain mapping, vol. 31, no. 8, pp. 1260--1271, January 2010. 
\href{https://doi.org/10.1002/hbm.20933}{https://doi.org/10.1002/hbm.20933}

\bibitem{Nunez_2019} P. L. Nunez, M. D. Nunez and R. Srinivasan,  ``Multi-scale neural sources of EEG: genuine, equivalent, and representative. A tutorial review,'' in Brain Topography, vol. 32, pp. 193--214, January 2019.  \href{https://doi.org/10.1007/s10548-019-00701-3}{https://doi.org/10.1007/s10548-019-00701-3}

\bibitem{Cisotto_2024} G. Cisotto and D. Chicco, ``Ten quick tips for clinical electroencephalographic (EEG) data acquisition and signal processing,'' in PeerJ Computer Science, vol. 10, pp. e2256, September 2024.  \href{http://dx.doi.org/10.7717/peerj-cs.2256}{http://dx.doi.org/10.7717/peerj-cs.2256}

\bibitem{eye_bci_multi_dataset} E. Guttmann-Flury, X. Sheng and X. Zhu, ``Eye-BCI multimodal dataset,'' Synapse 2024. Available: \href{https://doi.org/10.7303/syn64005218}{https://doi.org/10.7303/syn64005218}

\bibitem{GuttmannFlury_Dataset_2025} E. Guttmann-Flury, X. Sheng and X. Zhu, ``Dataset combining EEG, eye-tracking, and high-speed video for ocular activity analysis across BCI paradigms,'' in Sci. Data, vol. 12, no. 587, April 2025. \href{https://doi.org/10.1038/s41597-025-04861-9}{https://doi.org/10.1038/s41597-025-04861-9}

\bibitem{GuttmannFlury_APriori_2019} E. Guttmann-Flury, X. Sheng and X. Zhu, ``A priori sample size determination for the number of subjects in an EEG experiment,'' in 41st Annual International Conference of the IEEE Engineering in Medicine and Biology Conference (EMBC), Berlin, Germany, pp. 5180--5183, July 2019.	\href{https://doi.org/10.1109/EMBC.2019.8857482}{https://doi.org/10.1109/EMBC.2019.8857482}

\bibitem{GuttmannFlury_BadChannel_2025} E. Guttmann-Flury, W. Wei and S. Zhao, ``Automatic Blink-based Bad EEG channels Detection for BCI
Applications,'' in 47th Annual International Conference of the IEEE Engineering in Medicine and Biology Conference (EMBC), Copenhagen, Denmark, Accepted.	\href{https://doi.org/10.1016/j.compbiomed.2019.103442}{https://doi.org/10.1016/j.compbiomed.2019.103442}	

\bibitem{SuarezRevelo_2018} J. X. Suárez-Revelo, J. F. Ochoa-Gómez and C. A. Tobón-Quintero, ``Validation of EEG pre-processing pipeline by test-retest reliability,'' in Applied Computer Sciences in Engineering. WEA 2018. Communications in Computer and Information Science, vol 916. Springer, Cham, September 2018.	\href{https://doi.org/10.1007/978-3-030-00353-1_26}{https://doi.org/10.1007/978-3-030-00353-1\_26}	

\bibitem{Meng_2018} J. Meng, B. J. Edelman, J. Olsoe, G. Jacobs, S. Zhang, A. Beyko and B. He, ``A study of the effects of electrode number and decoding algorithm on online EEG-based BCI behavioral performance,'' in Frontiers in neuroscience, vol. 12, no. 227, April 2018.
\href{https://doi.org/10.3389/fnins.2018.00227}{https://doi.org/10.3389/fnins.2018.00227}

\bibitem{GuttmannFlury_Channel_2022} E. Guttmann-Flury, X. Sheng and X. Zhu, ``Channel selection from source localization: A review of four EEG-based brain–computer interfaces paradigms,'' in Behavior Research Methods, vol. 55, no. 4, pp. 1980--2003, July 2022.
\href{https://doi.org/10.3758/s13428-022-01897-2}{https://doi.org/10.3758/s13428-022-01897-2}

\bibitem{Hjorth_1991} B. Hjorth, ``Principles for transformation of scalp EEG from potential field into source distribution,'' in Journal of Clinical Neurophysiology, vol. 8, no. 4, pp. 391--396, October 1991.

\bibitem{Nunez_2006} P. L. Nunez and R. Srinivasan, ``Electric fields of the brain: the neurophysics of EEG,'' in Oxford university press, 2006.

\bibitem{Buxhoeveden_2002} D. P. Buxhoeveden and M. F. Casanova, ``The minicolumn hypothesis in neuroscience,'' in Brain, vol. 125, no. 5, pp. 935--951, May 2002.
\href{https://doi.org/10.1093/brain/awf110}{https://doi.org/10.1093/brain/awf110}	

\bibitem{Sporns_2007} O. Sporns, ``Brain connectivity,'' in Scholarpedia, vol. 2, no. 10, pp. 4695, 2007.

\bibitem{Gramfort_2013} A. Gramfort, M. Luessi, E. Larson, D. A. Engemann, D. Strohmeier, C. Brodbeck, R. Goj, M. Jas, T. Brooks, L. Parkkonen and M. S. Hämäläinen, ``MEG and EEG data analysis with MNE-Python,'' in Frontiers in Neuroscience, vol. 7, no. 267, pp.1 -- 13, December 2013.	\href{https://doi.org/10.3389/fnins.2013.00267}{https://doi.org/10.3389/fnins.2013.00267}

\bibitem{PascualMarqui_2018} R. D. Pascual-Marqui, P. Faber, T. Kinoshita, K. Kochi, P. Milz, K. Nishida and M. Yoshimura,  ``Comparing EEG/MEG neuroimaging methods based on localization error, false positive activity, and false positive connectivity,'' in BioRxiv, 269753, 2018.	\href{https://doi.org/10.1101/269753 }{https://doi.org/10.1101/269753}


\bibitem{Stam_2007} C. J. Stam, G. Nolte, and A. Daffertshofer, ``Phase lag index: assessment of functional connectivity from multi channel EEG and MEG with diminished bias from common sources,'' in Human Brain Mapping, vol. 28, no. 11, pp. 1178 -- 1193, January 2007.	\href{https://doi.org/10.1002/hbm.20346}{https://doi.org/10.1002/hbm.20346}

\bibitem{McFarland_2015} D. J. McFarland, ``The advantages of the surface Laplacian in brain–computer interface research,'' in International Journal of Psychophysiology, vol. 97, no. 3, pp.271--276, September 2015.	


\bibitem{Liu_2021} S. Liu, G. Li, S. Jiang, X. Wu, J. Hu, D. Zhang, and L. Chen, ``Investigating data cleaning methods to improve performance of brain–computer interfaces based on stereo-electroencephalography,'' Frontiers in Neuroscience, vol. 15, p.725384, June 2021.
\href{https://doi.org/10.3389/fnins.2021.725384}{https://doi.org/10.3389/fnins.2021.725384}






\end{thebibliography}
\end{document}